\RequirePackage{fixltx2e}
\documentclass[twocolumn,floatfix,showpacs,preprintnumbers,nofootinbib%,superscriptaddress
]{revtex4}
\usepackage[english]{babel}
\usepackage[utf8x]{inputenc}
\usepackage{mathrsfs}
\usepackage{amssymb}
\usepackage{amsmath}
\usepackage{mathtools}
\usepackage{mathrsfs}
\usepackage{graphicx}
\usepackage{fixltx2e}
\usepackage{tikz}

\usepackage{xcolor}
\definecolor{lcolor}{rgb}{0.5,0,0}
\definecolor{citcolor}{rgb}{0,0.3,0.0}

\usepackage[breaklinks,colorlinks,urlcolor=blue,citecolor=citcolor,linkcolor=lcolor]{hyperref}
\usepackage{subdepth}

\newcommand{\rt}{{\mathbf{r}}}
\newcommand{\xt}{{\mathbf{x}}}
\newcommand{\bt}{{\mathbf{b}}}

\newcommand{\yt}{{\mathbf{y}}}

\newcommand{\qt}{{\mathbf{q}}}
\newcommand{\kt}{{\mathbf{k}}}

\newcommand{\ptt}{p_\perp} % scalar
\newcommand{\ktt}{k_\perp} % scalar
\newcommand{\qtt}{q_\perp} % scalar
\newcommand{\ltt}{l_\perp} % scalar
\newcommand{\lt}{\mathbf{l}}

\newcommand{\ud}{\mathrm{d}}

\newcommand{\tr}{\, \mathrm{Tr} \, }

\newcommand{\nc}{{N_\mathrm{c}}}
\newcommand{\nf}{{n_\mathrm{f}}}

\newcommand{\cf}{C_\mathrm{F}}

\newcommand{\nr}[1]{(\ref{#1})}

\newcommand{\qso}{Q_\mathrm{s,0}}

\newcommand{\lqcd}{\Lambda_{\mathrm{QCD}}}
\newcommand{\as}{\alpha_{\mathrm{s}}}

\newcommand{\eq}{Eq.~}

\newcommand{\ical}{\mathcal{I}}
\newcommand{\jcal}{\mathcal{J}}
\newcommand{\scal}{\mathcal{S}}
\newcommand{\kcal}{\mathcal{K}}

\def\figscale{1.5}
\def\figspace{4mm}

\allowdisplaybreaks

\begin{document}

\author{B. Duclou\'e}
\author{T. Lappi}
\author{Y. Zhu}
\affiliation{
Department of Physics, % University of Jyv\"askyl\"a %
 P.O. Box 35, 40014 University of Jyv\"askyl\"a, Finland
}
\affiliation{
Helsinki Institute of Physics, P.O. Box 64, 00014 University of Helsinki,
Finland
}

\title{On the implementation of NLO  high energy factorization in single inclusive forward hadron production
}


\preprint{}

\begin{abstract}
Single inclusive particle production cross sections in high energy hadron collisions at forward rapidity are an important benchmark process for the CGC picture of small $x$ QCD. Recent calculations of this process have not led to a stable perturbative expansion for this quantity at high transverse momenta. We consider the quark channel production cross section using the new rapidity factorization procedure proposed by Iancu et al. We show that for fixed coupling one does indeed obtain a physically meaningful cross section which is positive and reduces in a controlled way to previous leading order calculations. We also consider a running coupling that depends on the transverse momentum of the produced particle. This gives a stable result which, however,  is not fully consistent with previous leading order calculations that use a coordinate space running coupling.
\end{abstract}

\maketitle

\section{Introduction}

To study the nonlinear gluon saturation regime one needs to look at processes that involve a momentum fraction $x$ in the target as small as possible, while still having a large enough transverse momentum to justify a weak coupling treatment. At LHC energies a good process to achieve this is particle production at forward rapidities. In the CGC picture this process can be calculated using the ``hybrid'' formalism, where a quark or a gluon, taken from the usual collinear parton distribution in the probe at large $x$, passes through the target color field. It receives a kick from the intrinsic transverse momentum of the target gluons, leading to the $\ptt$ spectrum of the produced hadrons.

Recently much work has been done to calculate forward particle production in the hybrid  picture to next-to-leading order (NLO) accuracy in the QCD coupling. Cross sections for this process were calculated to NLO accuracy in Refs.~\cite{Chirilli:2011km,Chirilli:2012jd} (see also the earlier works \cite{Dumitru:2005gt,Altinoluk:2011qy,JalilianMarian:2011dt}), where soft and collinear divergences in the one-loop calculation were factorized respectively into the BK~\cite{Balitsky:1995ub,Kovchegov:1999yj} evolution of the target and the DGLAP evolution of the probe proton and the fragmentation functions. In the first numerical implementation~\cite{Stasto:2013cha} of the ``CXY'' factorization framework of~\cite{Chirilli:2011km,Chirilli:2012jd} the NLO corrections turned out to be large and render the total cross section negative at large transverse momenta of the produced particle, signaling a problem in the organization of the perturbative series. Following this observation, several interpretations of the origin of the problem have been proposed \cite{Kang:2014lha,Altinoluk:2014eka,Watanabe:2015tja}. 

We recently argued~\cite{Ducloue:2016shw} that the origin of the problem lies in the subtraction procedure where the soft divergence in the NLO calculation is factorized into  the BK evolution of the target. Following this discussion, Iancu et al.~\cite{Iancu:2016vyg} suggested a new formulation of the NLO cross section in a way that explicitly yields a positive cross section. The formulation is based on the observation that in a certain limit which we will discuss in more detail below, the expression for the cross section involves similar structures as an integral form of the BK equation. If one takes care not to break this equivalence by further approximations and chooses a Fourier-positive initial condition for the BK equation, this will guarantee the positivity of the cross section. The resulting expression can either be written as a ``subtracted'' cross section, or in a form where no explicit subtraction is performed.

The purpose of this paper is to present a practical numerical implementation of the manifestly positive formulation for the cross section presented in \cite{Iancu:2016vyg}. We will first briefly review the expressions for the cross section in Sec.~\ref{sec:nlosinc}. We then in Sec.~\ref{sec:fc} explicitly show that this procedure works at fixed coupling as anticipated. 
In Sec.~\ref{sec:rc}  we discuss the problems associated with introducing a running coupling into the  expression. We implement the calculation using a running coupling that depends on the transverse momentum of the produced particle, which would be the natural thing to do for a cross section. We show that since the BK equation, and fits to DIS data using it, are usually implemented in coordinate space, this introduces a mismatch between the manifestly positive form of the cross section and a ``subtracted'' version where the separation between LO and NLO contributions is explicit. We then conclude in Sec.~\ref{sec:outlook} with a brief outlook for practical phenomenological applications of the formalism. In Appendix~\ref{sec:app_rc} we discuss an alternative coordinate space formulation of the cross section, which does not suffer from this mismatch between manifestly positive and subtracted formulations, but results in unphysically large values for the NLO corrections to the cross section. 

\section{Single inclusive particle production at NLO}
\label{sec:nlosinc}

Our starting point are the CXY formulae derived in Refs.~\cite{Chirilli:2011km,Chirilli:2012jd}. We will concentrate here on the quark channel, for which the leading order cross section
is proportional to the Fourier-transform of the dipole operator
\begin{eqnarray}
\scal(\ktt)=\scal(\ktt,\bt)=\int \ud^2\rt e^{-i\kt\cdot\rt} S(\rt),
\\ 
S(\rt=\xt-\yt)=\left< \frac{1}{\nc}\tr V(\xt)V^\dag(\yt) \right>,
\end{eqnarray}
where $V(\xt)$ is a fundamental representation Wilson line in the color field of the target.
The NLO cross section involves also adjoint representation Wilson lines from the gluon interacting with the target color field. Using Fierz identities and the mean field approximation, which replaces expectation value of products of dipole operators by products of expectation values, also the NLO cross section can be expressed in terms of $\scal(\ktt)$. Since our goal is to study the negativity problem and how it would be affected by the proposal of Ref.~\cite{Iancu:2016vyg}, we leave out the fragmentation functions which do not play any role here. Following the notation in \cite{Ducloue:2016shw}, the (unsubtracted) CXY quark multiplicity can be written as:
\begin{widetext}
\begin{align}\label{eq:nlosigma}
 \frac{\ud N^{pA \to qX}}{\ud^2\kt \ud y}
=
x_p q(x_p) \frac{\scal_0(\ktt) }{(2\pi)^2} &+ \frac{\as}{2\pi^2}
\int_{x_p}^{\xi_\text{max}} \ud \xi \frac{1+\xi^2}{1-\xi}
\frac{x_p}{\xi} q\left(\frac{x_p}{\xi}\right) \left\{\cf \ical(\ktt,\xi,X(\xi)) + \frac{\nc}{2}\jcal(\ktt,\xi,X(\xi)) \right\} \nonumber \\
&- \frac{\as}{2\pi^2}
\int_{0}^{\xi_\text{max}} \ud \xi \frac{1+\xi^2}{1-\xi}
x_p q\left(x_p \right) \left\{\cf \ical_v(\ktt,\xi,X(\xi)) + \frac{\nc}{2}\jcal_v(\ktt,\xi,X(\xi)) \right\} .
\end{align}
\end{widetext}
Here we have left $\xi_\text{max}$, the upper limit for the $\xi$-integrals, unspecified for now as we will return to this question shortly.
The kinematical variables are defined as
$x_p=\ktt e^{y}/\sqrt{s}$, $x_g=\ktt e^{-y}/\sqrt{s}$ and  $\ktt=|\kt|$. The most important one for our discussion here is the momentum fraction $\xi$: the fragmenting quark carries a fraction $\xi$ of the incoming quark longitudinal momentum. Thus the incoming quark has a momentum fraction $x_p/\xi$ of the incoming proton, where $x_p$ is the probe momentum fraction in the leading order kinematics.  The radiated gluon in the NLO terms carries a longitudinal momentum fraction $1-\xi$, i.e. the limit $\xi\to 1$ corresponds to the soft gluon emission that  must be resummed into the BK evolution of the target. For the following discussion it is important to note the interpretation of the variable $x_g$: it is easy to see that for producing a final state quark with transverse momentum $k_\perp$ at leading order, $x_g$ is the fraction of target longitudinal momentum $P^-$ needed to put this quark (with longitudinal momentum $x_p P^+$) on shell.
Thus, in a  leading order calculation such as \cite{Lappi:2013zma}, with BK evolution starting from an initial momentum fraction $x_0\sim 0.01$, one would evolve the target for $\ln x_0/x_g$ units in rapidity (here rapidity is defined as the logarithm of the inverse of the target momentum fraction).

The dependence on the Wilson line correlators in the target comes through transverse momentum integrals that can be expressed in terms of the dipole operator $\scal$:
\begin{widetext}
\begin{eqnarray}
\ical(\ktt,\xi,X(\xi)) \!\! &=& \!\!
\int \frac{\ud^2 \qt}{(2\pi)^2} 
\left[\frac{\kt-\qt}{(\kt-\qt)^2} - \frac{\kt-\xi \qt}{(\kt-\xi \qt)^2} \right]^2
\scal(\qtt,X(\xi)) \, ,
\\ 
% ******************
\label{eq:defJ}
\jcal(\ktt,\xi,X(\xi)) \!\! &=& \!\!
\int \frac{\ud^2 \qt}{(2\pi)^2} 
\frac{2(\kt-\xi\qt)\cdot(\kt-\qt)}{(\kt-\xi\qt)^2(\kt-\qt)^2}
\scal(\qtt,X(\xi))
\\ &&
-\int \frac{\ud^2 \qt}{(2\pi)^2} \frac{ \ud^2\lt}{(2\pi)^2}
\frac{2(\kt-\xi\qt)\cdot(\kt-\lt)}{(\kt-\xi\qt)^2(\kt-\lt)^2}
\scal(\qtt,X(\xi))\scal(\ltt,X(\xi)) \, ,
\\
% ******************
\ical_v(\ktt,\xi,X(\xi)) \!\! &=& \!\!
\int \frac{ \ud^2 \qt }{(2\pi)^2}
\left[\frac{\kt-\qt}{(\kt-\qt)^2} - \frac{\xi\kt-\qt}{(\xi \kt-\qt)^2} \right]^2
\scal(\ktt,X(\xi)) \, ,
\\
% ******************
\label{eq:defJv}
\jcal_v(\ktt,\xi,X(\xi)) \!\! &=& \!\!
\left[
\int \frac{\ud^2 \qt}{(2\pi)^2} 
\frac{2(\xi\kt-\qt)\cdot(\kt-\qt)}{(\xi\kt-\qt)^2(\kt-\qt)^2}
-\int \frac{\ud^2 \qt}{(2\pi)^2} \frac{  \ud^2\lt}{(2\pi)^2}
\frac{2(\xi\kt-\qt)\cdot(\lt-\qt)}{(\xi\kt-\qt)^2(\lt-\qt)^2}
\scal(\ltt,X(\xi))
\right]
\scal(\ktt,X(\xi))
.
\end{eqnarray}
\end{widetext}
At this stage we have not specified how the dipole operators depend  on the energy or rapidity, but have just denoted this dependence by $X(\xi)$. This dependence is indeed the crux of the argument in Ref.~\cite{Iancu:2016vyg}, which we will here reproduce in a slightly altered form. As emphasized e.g. in \cite{Altinoluk:2014eka,Ducloue:2016shw}, one achieves a more stable perturbative expansion when this energy dependence is related to the momentum fraction in the target, i.e. the $k^-$ scale in the scattering. This distinction becomes important at large transverse momenta for the produced particle~\cite{Kang:2014lha,Altinoluk:2014eka,Watanabe:2015tja,Ducloue:2016shw}. However, the formulation of 
Ref.~\cite{Iancu:2016vyg} allows one to get a physically reasonable cross section even without imposing a separate ``kinematical constraint,''  within just the usual ``Regge'' kinematics, where all transverse momenta are assumed to be of the same order. Thus we will for now keep this assumption, which allows us to relate the momentum fraction in the target to the kinematics of the probe as $X(\xi)\approx x_g/(1-\xi)$.

The CXY cross sections have been derived using the eikonal approximation, which allows one to describe the interaction with the target in terms of Wilson lines. This approximation is only valid for sufficiently high energy scattering, where now ``high energy'' should refer to the whole quark-gluon state. This is usually reflected in a restriction that the momentum fraction $X(\xi)$ at which one evaluates the dipole cross sections must be smaller than some limiting value $x_0\sim 0.01$. For high energy evolution this corresponds to a nonperturbative initial condition for the evolution of the target which is fit to experimental data. With $X(\xi)=x_g/(1-\xi)$ this means that the $\xi$-integration must be restricted to $\xi<1-x_g/x_0\equiv \xi_\text{max}$. This now defines the upper limit of the integration that was left unspecified in \eq\nr{eq:nlosigma}.  Note that this restriction implies that extrapolating beyond the physical region $\xi_\text{max}> x_p$ to the kinematical limit $x_g=x_0$, the NLO corrections are explicitly set to approach zero. Calculating NLO corrections involving larger target longitudinal momenta $k^-$ is strictly speaking not possible in this formalism and would require going beyond the eikonal approximation. We would argue that setting these contributions to zero is a more controlled approximation than letting them evaluate to some arbitrary value. This should be kept in mind when comparing these calculations to experimental data: this restriction means that when one approaches the kinematical limit $x_g \to x_0$, the phase space for NLO contributions in our calculation is cut off by this constraint. In this limit the calculation could eventually be matched onto collinear factorization, which becomes more appropriate when $x_g$ is large, as done in Ref.~\cite{Stasto:2014sea}.

\subsection{$\nc$-terms}

In Ref.~\cite{Ducloue:2016shw} it was shown that the negativity of the CXY cross section at large transverse momentum is caused by the NLO corrections proportional to $\nc$, which become very large and negative while the NLO corrections proportional to $\cf$ are positive. Therefore we first consider only the leading order and $\nc$-terms, as was done in Ref.~\cite{Iancu:2016vyg}. These contributions can be written as
\begin{align}\label{eq:nc_init}
\frac{\ud N^{\text{LO}+\nc}}{\ud^2\kt \ud y}
=& x_p q(x_p) \frac{\scal_0(\ktt)}{(2\pi)^2} \nonumber \\
&+ \as
\int_0^{1-x_g/x_0} \frac{\ud \xi}{1-\xi} \kcal(\ktt,\xi,X(\xi)) ,
\end{align}
where $\kcal$ is defined as
\begin{align}
\kcal(\ktt,\xi,X(\xi))=&\frac{\nc}{(2\pi)^2}(1+\xi^2) \nonumber \\
\times \bigg[&\theta(\xi-x_p)\frac{x_p}{\xi} q\left(\frac{x_p}{\xi}\right) \jcal(\ktt,\xi,X(\xi)) \nonumber \\
&-x_p q\left(x_p \right)
\jcal_v(\ktt,\xi,X(\xi)) \bigg].
\end{align}
At this stage the dipole operator $\scal_0$ is a ``bare'' one and does not evolve with rapidity. Since $\kcal(\ktt,\xi,X)$ approaches a nonzero value for $\xi \to 1 $ at fixed $X$, the cross section contains a large logarithmic integral in the high energy limit $x_g \to 0$; this should be resummed using the BK equation. The ``bare'' dipole $\scal_0$ is then identified with the initial condition of the BK evolution, formulated at a given initial rapidity $\ln\frac{1}{x_0}$. We thus rewrite~(\ref{eq:nc_init}) as
\begin{align}\label{eq:nc_unsub}
\frac{\ud N^{\text{LO}+\nc}}{\ud^2\kt \ud y}
=& x_p q(x_p) \frac{\scal(\ktt,x_0)}{(2\pi)^2} \nonumber \\
&+ \as
\int_0^{1-x_g/x_0} \frac{\ud \xi}{1-\xi} \kcal(\ktt,\xi,X(\xi)) \nonumber \\
\equiv & \frac{\ud N^{\text{IC}}}{\ud^2\kt \ud y}+ \frac{\ud N^{\nc, unsub}}{\ud^2\kt \ud y} \, ,
\end{align}
where ``IC'' stands for the multiplicity at the initial rapidity scale $x_0$.
As long as the initial dipole amplitude is Fourier-positive (which is a nontrivial requirement~\cite{Giraud:2016lgg} that we will however assume to be satisfied here, as it is for the parametrizations that we will use), the multiplicity~(\ref{eq:nc_unsub}) is positive up to large transverse momenta. 
Now, since we are taking   $X(\xi) = x_g/(1-\xi)$  as is appropriate in the Regge kinematics, we have $\ud \ln X(\xi) = - \ud \ln (1-\xi)$. This enables us to write  an integral version of the BK equation as
\begin{align}\label{eq:integralBK}
&\scal(\ktt,x_g)=\scal(\ktt,x_0) \nonumber \\
&+2 \as \nc \!\!\!\! \int\limits_{0}^{1-x_g/x_0} \!\! \frac{\ud \xi}{1-\xi} 
\left[\jcal(\ktt,1,X(\xi)) - \jcal_v(\ktt,1,X(\xi))\right],
\end{align}
and to write, assuming that $x_p<1-x_g/x_0$ ,
\begin{multline}\label{eq:nc_sub}
\frac{\ud N^{\text{LO}+\nc}}{\ud^2\kt \ud y}
= x_p q(x_p) \frac{\scal(\ktt,x_g)}{(2\pi)^2} \\
+\as \!\!\!\!
\int\limits_0^{1-x_g/x_0} \!\! \frac{\ud \xi}{1-\xi} \left[\kcal(\ktt,\xi,X(\xi)) - \kcal(\ktt,1,X(\xi))\right]  \\
\equiv  \frac{\ud N^\text{LO}}{\ud^2\kt \ud y}+\frac{\ud N^{\nc,sub}}{\ud^2\kt \ud y} \, .
\end{multline}
Since this equation is equivalent to \nr{eq:nc_unsub}, it is positive even up to large transverse momenta. Equation~\nr{eq:nc_sub} is written explicitly as a sum of a leading order contribution (with a BK-evolved dipole) and a contribution proportional to $\as$ that has no large energy logarithm because $\kcal(\ktt,\xi,X(\xi)) - \kcal(\ktt,1,X(\xi))$ vanishes for $\xi \to1 $. Thus it can be naturally interpreted as the sum of a LO contribution and a NLO correction. In the version \nr{eq:nc_unsub}, on the other hand, the way to reduce the NLO expression to the LO limit is less transparent. Instead of dropping out a term explicitly proportional to $\as$, the LO limit of \nr{eq:nc_unsub} is taken by replacing $\kcal(\ktt,\xi,X(\xi))$ with $\kcal(\ktt,1,X(\xi))$, i.e. setting $\xi=1$ inside the kernel, but without changing the rapidity scale of the dipole correlators.

To go from  \nr{eq:nc_unsub} or~\nr{eq:nc_sub} to the CXY expressions one starts by noting that, because of the subtraction $\kcal(\ktt,\xi,X(\xi))-\kcal(\ktt,1,X(\xi))$, the integral~\nr{eq:nc_sub} is dominated by the lower limit $\xi\ll 1$. Thus one could argue that the rapidity of the dipole amplitude can be  replaced by its value at $\xi=0$,  i.e.  $\scal(\ktt,X(\xi))$ by $\scal(\ktt,X(0))=\scal(\ktt,x_g)$ and similarly for $\scal(\qtt,X(\xi))$ and $\scal(\ltt,X(\xi))$. This approximation
\begin{multline}\label{eq:cxyapprox1}
\frac{\ud N^{\text{LO}+\nc}}{\ud^2\kt \ud y}
= x_p q(x_p) \frac{\scal(\ktt,x_g)}{(2\pi)^2} \\
+ \as
\int_0^{1-x_g/x_0} \frac{\ud \xi}{1-\xi} \left[\kcal(\ktt,\xi,x_g)-\kcal(\ktt,1,x_g)\right],
\end{multline}
is perfectly justified in a weak coupling sense. However, as pointed out in  \cite{Ducloue:2016shw,Iancu:2016vyg}, it  makes the cross section negative, because at large transverse momentum the subtracted NLO term $\kcal(\ktt,\xi,x_g)-\kcal(\ktt,1,x_g)$ is negative and can dominate over the leading order term. Thus the necessary  ingredient in keeping the result physically meaningful turns out to be to not just evaluate the dipole amplitude at the rapidity scale $x_g$ of the leading order cross section, but to keep the dependence of $X(\xi)$ on the integration variable.

In addition to changing the rapidity argument of the dipole operators, another approximation is needed to recover the CXY subtraction result, namely  to  replace $1-x_g/x_0$ in the upper limit of the integration over $\xi$  in \eq\nr{eq:cxyapprox1} by 1. Now that the rapidity argument $X(\xi)$ in the dipole operators has been replaced by $x_g$, changing the integration limit is formally possible without extending the dipole parametrization to the large $x$ region where the eikonal approximation is not valid. But this would make the problem disappear only superficially: any contribution  from $\xi > 1-x_g/x_0$ would come from the region where the derivation of the original expression~\nr{eq:nlosigma} is dubious because of the large invariant mass of the  produced quark-gluon system. The question of the proper value of the upper limit in $\xi$, or of the correct dependence $X(\xi),$ for large transverse momenta of the produced particle is the ``kinematical constraint'' or ``Ioffe time'' issue that has been extensively discussed by several authors \cite{Altinoluk:2014eka,Watanabe:2015tja,Ducloue:2016shw}. We will however not attempt to take these corrections into account here, but stay within the Regge kinematical approximation $X(\xi) = x_g/(1-\xi)$. 

In the rest of this paper we will refer to the explicitly positive formulation \nr{eq:nc_unsub} as the ``unsubtracted'' and its subtracted version~\nr{eq:nc_sub} as the ``subtracted'' formulation.

\subsection{$\cf$-terms}

Let us now turn to the $\cf$-terms. While they do not pose similar manifest problems as the $\nc$-terms, one nevertheless has to make certain choices concerning the rapidity dependence of the dipole correlators when evaluating the cross section.
Writing explicitly the integration limits and subtracting the $1/\varepsilon$ poles corresponding to the DGLAP evolution of the probe quark distributions and fragmentation functions (which are not written explicitly here), the $\cf$-terms become
\begin{widetext}
\begin{equation}\label{eq:cf}
\frac{\ud N^{\cf}}{\ud^2\kt \ud y} \equiv \frac{\as}{2\pi^2} \cf
\left[\int\limits_{x_p}^{1-x_g/x_0} \ud \xi \frac{1+\xi^2}{1-\xi} 
\frac{x_p}{\xi} q\left(\frac{x_p}{\xi}\right) \ical^\text{finite}(\ktt,\xi,X(\xi)) 
- \int\limits_{0}^{1-x_g/x_0} \ud \xi \frac{1+\xi^2}{1-\xi}
x_p q\left(x_p \right) \ical_v^\text{finite}(\ktt,\xi,X(\xi)) \right] ,
\end{equation}
where $\ical^\text{finite}$ and $\ical_v^\text{finite}$, obtained after subtracting the collinear divergence from $\ical$ and $\ical_v$, read
\begin{align}
\label{eq:Ifinite}
\ical^\text{finite}(\ktt,\xi,X(\xi)) &= \int \frac{\ud^2\rt}{4\pi} S(\rt,X(\xi)) \ln \frac{c_{0}^{2}}{\rt^{2}\mu ^{2}}\left( e^{-i\kt\cdot \rt}+\frac{1}{\xi^2}e^{-i\frac{\kt}{\xi }\cdot \rt}\right)
- 2 \int \frac{\ud^2 \qt}{(2\pi)^2} \frac{(\kt-\xi\qt)\cdot(\kt-\qt)}{(\kt-\xi\qt)^2(\kt-\qt)^2} \scal(\qtt,X(\xi)) \, ,\\
\label{eq:Ivfinite}
\ical_v^\text{finite}(\ktt,\xi,X(\xi)) &= \frac{\scal(\ktt,X(\xi))}{2\pi} \left(\ln{\frac{\ktt^2}{\mu^2}}+ \ln(1-\xi)^2\right).
\end{align}
\end{widetext}
To arrive at the expressions~(\ref{eq:cf})-(\ref{eq:Ivfinite}) we have adopted a specific choice for the previously mentioned ambiguity: namely the rapidity scale at which the dipole operators are evaluated in the $\cf$-terms. To be more precise: to get a finite result for the $\cf$-terms one must subtract from the cross section collinear infrared divergences that manifest themselves as $1/\varepsilon$ poles  in the cross section. Thus one is subtracting terms that  are proportional to the bare dipole operator $\scal_0$ from the NLO cross section \nr{eq:nlosigma} and neglecting a collinearly divergent remainder $ \sim \as \cf (1/\varepsilon) \left(\scal(X(\xi)) - \scal_0\right)$. 
Such terms are formally of NNLO order, since the difference $\left(\scal(X(\xi)) - \scal(x_0)\right)$ is proportional to $\as$. This difference can, however, be numerically large, indeed the large difference between $\scal(X(\xi))$, $\scal(x_g)$ and $\scal_0$ was essential for understanding the negativity of the $\nc$-terms. 
In the CXY approximation  these terms turn into $\sim \as \cf (1/\varepsilon) \left(\scal(x_g) - \scal_0\right)$,
which can similarly be neglected at this level of accuracy.
%At this level of accuracy we could have replaced $X(\xi)$ by $x_g$ in the $\cf$-term. 
For the $\nc$-terms the choice between $X(\xi)$ and $x_g$ was dictated by the desire to maintain the relation between the integral form of the BK equation and the subtraction. For the $\cf$-terms there is no such requirement. Our choice here is to  keep the same energy dependence $X(\xi)$, and the same kinematical limit $\xi<1-x_g/x_0$ for the $\cf$-terms as for the $\nc$-terms. 

Because $\ical$ and $\ical_v$ vanish at $\xi=1$, replacing the upper limit 
$\xi < 1-x_g/x_0$ by $\xi<1$ would be a rather good approximation here as long as we are safely inside the domain of validity of the eikonal approximation $x_g \ll x_0$. Evaluating the dipoles at different rapidity arguments $X(\xi)$ vs. $x_g$ makes a larger difference. However also here the difference becomes smaller at smaller $x_g$, because the integral over $\xi$ is increasingly dominated by smaller $\xi$ where $X(\xi)\approx x_g$. The extremely forward kinematical limit $x_p \to 1$ is an exception to this argument, because the vanishing of the collinear quark distribution at $x_p/\xi =1$ suppresses the small $\xi$ contribution to the cross section, making the result more sensitive to the difference between $X(\xi)$ and $x_g$.
We will demonstrate the effect of these different choices for the $\cf$-terms in a particular kinematical configuration in the next section.

\begin{figure*}
	\includegraphics[scale=\figscale]{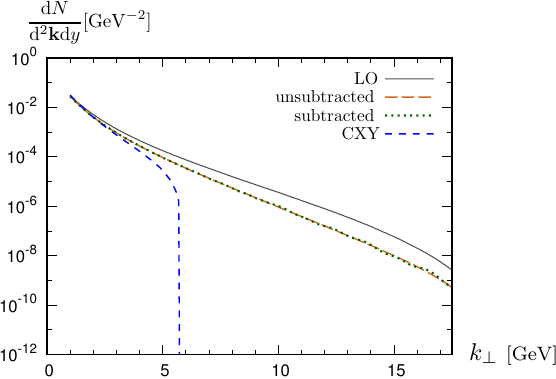}
	\hspace{\figspace}
	\includegraphics[scale=\figscale]{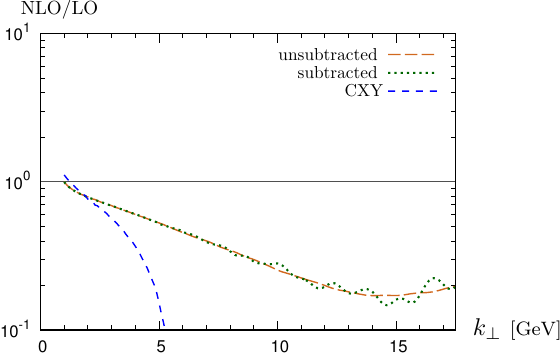}
	\caption{The LO and $\nc$-parts of the quark production multiplicity calculated with a fixed coupling for $\sqrt{s}=500$ GeV and at rapidity $y=3.2$, showing the NLO result using the unsubtracted~(\ref{eq:nc_unsub}) and subtracted~(\ref{eq:nc_sub}) expressions and the CXY approximation which replaces the upper limit of the $\xi$ integration by $1$ and evaluates the dipole correlators at $X(\xi)=x_g$. On the left the multiplicity and on the right the ratio to the LO result.}
	\label{fig:fcBK_Nc}
\end{figure*}

\begin{figure*}
	\includegraphics[scale=\figscale]{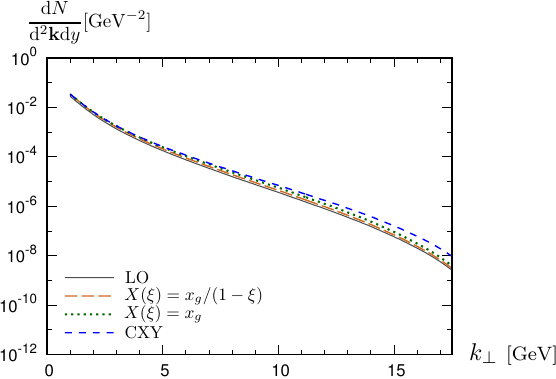}
	\hspace{\figspace}
	\includegraphics[scale=\figscale]{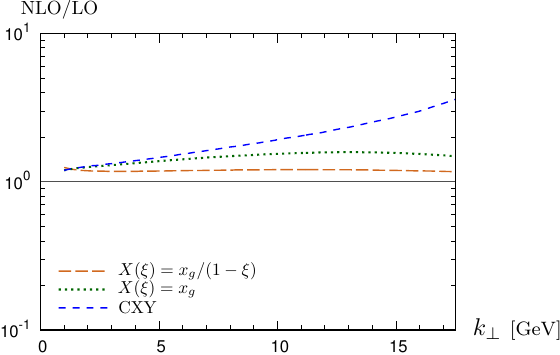}
	\caption{The LO and $\cf$-parts of the quark production multiplicity calculated with a fixed coupling for $\sqrt{s}=500$ GeV and at rapidity $y=3.2$, showing the NLO result~(\ref{eq:sigma_cf}) evaluated either with $X(\xi)=x_g$ or $X(\xi)=x_g/(1-\xi)$ and the CXY approximation which replaces the upper limit of the $\xi$ integration by $1$ and evaluates the dipole correlators at $X(\xi)=x_g$. On the left the multiplicity and on the right the ratio to the LO result.}
	\label{fig:fcBK_CF}
\end{figure*}

\begin{figure*}
	\includegraphics[scale=\figscale]{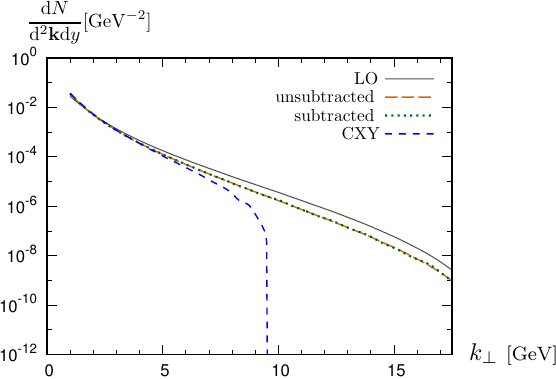}
	\hspace{\figspace}
	\includegraphics[scale=\figscale]{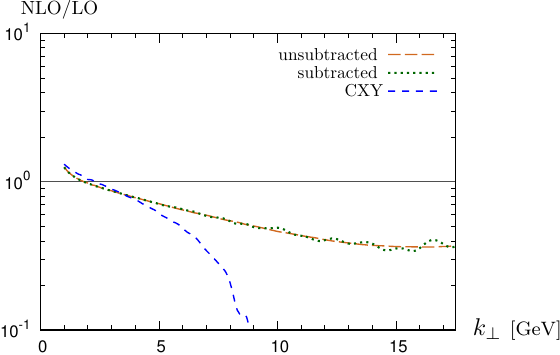}
	\caption{The complete NLO quark channel multiplicity including both the $\nc$ and $\cf$ corrections calculated with a fixed coupling for $\sqrt{s}=500$ GeV and at rapidity $y=3.2$, showing the NLO result using the unsubtracted~(\ref{eq:NLO_unsub}) and subtracted~(\ref{eq:NLO_sub}) expressions and the CXY approximation which replaces the upper limit of the $\xi$ integration by $1$ and evaluates the dipole correlators at $X(\xi)=x_g$. On the left the multiplicity and on the right the ratio to the LO result.}
	\label{fig:fcBK_NLO}
\end{figure*}

\section{Numerical results: fixed coupling}
\label{sec:fc}

%\section{Numerical results}

We now turn to  numerical results for the quark production cross section. In practice we consider up quark production and use the MSTW2008 NLO parametrization~\cite{Martin:2009iq} for the quark distribution $q(x)$ in the projectile. This distribution is evaluated at the transverse scale $Q=\ktt$. The center of mass energy of the collision and the rapidity of the produced quark are chosen as $\sqrt{s}=500$ GeV and $y=3.2$ respectively. In these kinematics, the limit $x_p=1-x_g/x_0$ is reached when $\ktt \sim 17.5$ GeV for $x_0=0.01$. As explained previously, we do not consider our calculation to be reliable for transverse momenta larger than this value so we will not show results above this limit.

We first consider a fixed value of the strong coupling, $\as=0.2$. The rapidity dependence of the dipole operator of the target is obtained by solving numerically the LO BK equation with $\as=0.2$ and an MV initial condition~\cite{McLerran:1993ni},
\begin{equation}
	S(\rt,x_0)=\exp\left[-\frac{\rt^2 \qso^2}{4}\ln{\left(\frac{1}{|\rt| \lqcd}+e\right)}\right],
\end{equation}
with $\qso^2=0.2$ GeV$^2$ and $\lqcd=0.241$ GeV. Combining Eqs.~(\ref{eq:nc_unsub}) and~(\ref{eq:cf}), the NLO multiplicity reads
\begin{equation}\label{eq:NLO_unsub}
\frac{\ud N^\text{NLO}}{\ud^2\kt \ud y}=\frac{\ud N^{\text{IC}}}{\ud^2\kt \ud y}
+ \frac{\ud N^{\nc, unsub}}{\ud^2\kt \ud y}
+ \frac{\ud N^{\cf}}{\ud^2\kt \ud y} .
\end{equation}
As discussed above, recovering the LO BK result from this formulation is not a matter of dropping terms that are explicitly proportional to $\as$. Instead this is done by setting $\xi=1$ everywhere in the integrand, except in the $\ud \xi/(1-\xi)$ term that gives the leading logarithmic contribution, and in $X(\xi)$ which is needed to recover the BK equation in integral form. The separation between LO and NLO terms becomes more explicit if we use the equivalence between~(\ref{eq:nc_unsub}) and~(\ref{eq:nc_sub}) to rewrite~\nr{eq:NLO_unsub} as
\begin{equation}\label{eq:NLO_sub}
\frac{\ud N^\text{NLO}}{\ud^2\kt \ud y}=\frac{\ud N^{\text{LO}}}{\ud^2\kt \ud y}
+ \frac{\ud N^{\nc, sub}}{\ud^2\kt \ud y}
+ \frac{\ud N^{\cf}}{\ud^2\kt \ud y} .
\end{equation}
This last expression is very similar to the subtracted CXY cross section, except for the rapidity scale at which the dipole operators in the NLO corrections are evaluated and the upper limit on the $\xi$-integrals. We will see in the following that these changes lead to very different numerical results at large $\ktt$ compared to the CXY expressions.

The fact that Eqs.~(\ref{eq:nc_unsub}) and~(\ref{eq:nc_sub}) are equivalent is demonstrated explicitly in Fig.~\ref{fig:fcBK_Nc}, where we show the LO+$\nc$ result obtained with these two equations. We observe that the results obtained with~(\ref{eq:nc_unsub}) are more stable numerically at large transverse momentum than when using~(\ref{eq:nc_sub}). This was to be expected since~(\ref{eq:nc_sub}) involves a cancellation between rather large contributions. One can also note that the NLO corrections proportional to $\nc$ are rather large and negative at large $\ktt$ but the LO+$\nc$ multiplicity is positive at all transverse momenta. This is in contrast with the CXY result, also shown on this figure, which becomes negative around $\ktt \sim 6$ GeV.

Let us now consider the LO+$\cf$ result,
\begin{equation}\label{eq:sigma_cf}
\frac{\ud N^{\text{LO}+\cf}}{\ud^2\kt \ud y}=\frac{\ud N^{\text{LO}}}{\ud^2\kt \ud y}
+ \frac{\ud N^{\cf}}{\ud^2\kt \ud y} \, .
\end{equation}
As already mentioned, the rapidity scale at which the dipole correlators are evaluated in the $\cf$-terms is left unspecified by the proposal~\cite{Iancu:2016vyg}, and the difference between evaluating these correlators at $X(\xi)$ or $X(0)=x_g$ is formally a higher order effect. In Fig.~\ref{fig:fcBK_CF} we show how the LO+$\cf$ result is affected by these different choices. Note that at similar rapidities and transverse momenta at LHC energies $x_g$ (and $x_p$) would be smaller, and the difference between these choices smaller. In the following we choose to evaluate the dipole correlators of the $\cf$-terms at $X(\xi)$ since it would be quite unnatural to use different rapidity scales in the $\cf$ and $\nc$-terms.

Now that we have specified how to evaluate both the $\nc$ and $\cf$-terms, we can sum these contributions to get the full NLO result. Since both the LO+$\nc$ result and the NLO $\cf$ corrections are positive at all $\ktt$, as shown in Figs.~\ref{fig:fcBK_Nc} and~\ref{fig:fcBK_CF} respectively, the NLO multiplicity is positive as well. This is shown in Fig.~\ref{fig:fcBK_NLO}, where one can observe that the net result of adding the $\nc$ and $\cf$ NLO corrections makes the cross section significantly smaller at high $\ktt$ than the leading order result. For comparison we also show the corresponding CXY result which becomes negative above $\ktt \sim 10$ GeV.

\section{Numerical results: running coupling}
\label{sec:rc}

\begin{figure}
	\includegraphics[scale=\figscale]{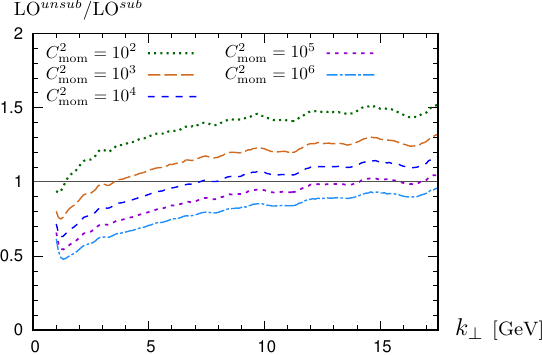}
	\caption{Ratio of the LO limits of Eq.~(\ref{eq:nc_unsub}), involving a momentum space running coupling~(\ref{eq:as_mom}), and Eq.~(\ref{eq:nc_sub}) for $\sqrt{s}=500$ GeV and at rapidity $y=3.2$, using different values for the parameter $C_\text{mom}^2$.}
	\label{fig:rcBK_LO}
\end{figure}

\begin{figure*}
	\includegraphics[scale=\figscale]{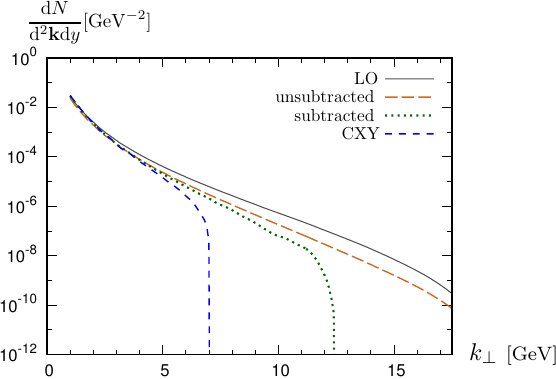}
	\hspace{\figspace}
	\includegraphics[scale=\figscale]{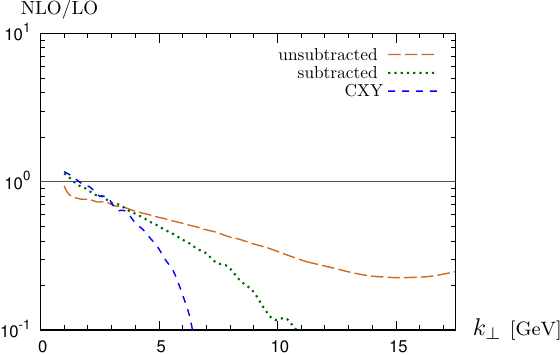}
	\caption{The complete NLO quark channel multiplicity including both the $\nc$ and $\cf$ corrections calculated with a momentum running coupling~(\ref{eq:as_mom}) with $C_\text{mom}^2=10^3$ for $\sqrt{s}=500$ GeV and at rapidity $y=3.2$, showing the NLO result using the unsubtracted~(\ref{eq:NLO_unsub}) and subtracted~(\ref{eq:NLO_sub}) expressions and the CXY approximation which replaces the upper limit of the $\xi$ integration by $1$ and evaluates the dipole correlators at $X(\xi)=x_g$. On the left the multiplicity and on the right the ratio to the LO result.}
	\label{fig:rcBK_NLO_mom}
\end{figure*}

A realistic QCD calculation must include the running of the coupling constant as a function of the scale. In the BK evolution equation this is typically done using the ``Balitsky'' prescription~\cite{Balitsky:2006wa}, as is the case for the dipole operators from~\cite{Lappi:2013zma} that we will be using here. Other choices are also possible (see e.g.~\cite{Albacete:2004gw,Kovchegov:2006vj,Iancu:2015joa}), but the difference between these is not essential for our discussion here. What is important is that the BK equation is normally solved in coordinate space, and therefore also the running coupling prescription involves a kernel with a coupling constant that depends on coordinate differences. While momentum space solutions are also possible, there are several reasons why the coordinate space formulation is preferable. Firstly, the initial condition for the evolution needs to be extracted from a fit to experimental data, typically from DIS, where the cross section is most naturally expressed in terms of the coordinate space dipole. The fitting procedure is thus most naturally done in coordinate space. Secondly, a fundamental property of the BK equation is  gluon saturation, which is guaranteed by the \emph{coordinate space} dipole amplitude $S(\rt)$ taking values between $0$ and $1$. In a coordinate space calculation this unitarity requirement is explicitly satisfied also for running coupling. On the other hand, momentum space running coupling versions of the equation, at least known ones, are equivalent to a coordinate space version only parametrically, but not exactly. They do not therefore automatically enforce unitarity. Because of these reasons we will also here take the stand that the BK evolution that one would prefer to use (in our case that of~\cite{Lappi:2013zma}) in conjunction with the cross section calculation will be a coordinate space running coupling one.

Parametrically the scale of the coupling constant is set in this case by the transverse momentum of the produced particle. The procedure suggested in Ref.~\cite{Iancu:2016vyg} is to simply replace the explicit coupling constant $\as$ appearing in Eqs.~(\ref{eq:nc_unsub}), (\ref{eq:nc_sub}) and~(\ref{eq:cf}) by $\as(\ktt^2)$. However, doing this while using dipole operators $\scal$ whose energy dependence results from a BK equation with  coordinate space running coupling spoils the exact equivalence between the manifestly positive \nr{eq:NLO_unsub} and subtracted \nr{eq:NLO_sub} expressions for the cross section. This mismatch can be studied by comparing the LO limits of Eqs.~(\ref{eq:nc_unsub}) and~(\ref{eq:nc_sub}). For this we use dipole correlators obtained by solving numerically the LO BK equation with the Balitsky prescription for the running coupling corrections~\cite{Balitsky:2006wa}. The initial condition is the MV$^e$ parametrization introduced in~\cite{Lappi:2013zma}:
\begin{equation}
S(\rt,x_0)=\exp\left[-\frac{\rt^2 \qso^2}{4}\ln{\left(\frac{1}{|\rt| \lqcd}+e_c \cdot e\right)}\right],
\end{equation}
and the running coupling in coordinate space is taken as
\begin{equation}\label{eq:as_pos}
\as(\rt^2)=\frac{4\pi}{\beta_0 \ln\left(\frac{4 C^2}{\rt^2 \lqcd^2}\right)},
\end{equation}
with $\beta_0=(11\nc-2\nf)/3$ and $\lqcd=0.241$ GeV. At large $|\rt|$ the coupling is frozen at the value 0.7. The values of the parameters in these expressions were obtained in~\cite{Lappi:2013zma} by a fit to HERA DIS data~\cite{Aaron:2009aa} as $\qso^2=0.06$ GeV$^2$, $C^2=7.2$ and $e_c=18.9$. This parametrization describes a target proton and not a nucleus, but this doesn't affect our discussion. More importantly, the MV$^e$ parametrization has a positive Fourier transform.

In Fig.~\ref{fig:rcBK_LO} we show the ratio between the LO limits of Eqs.~(\ref{eq:nc_unsub}) (involving a momentum space running coupling) and~(\ref{eq:nc_sub}) (BK with ``Balitsky''-coordinate space $\as$). Here we use the expression 
\begin{equation}\label{eq:as_mom}
\as(\ktt^2)=\frac{4\pi}{\beta_0 \ln\left(\frac{C_\text{mom}^2 \ktt^2}{\lqcd^2}\right)}
\end{equation}
for the momentum space running coupling that appears explicitly in the equations for the cross section, using different values of the parameter $C_\text{mom}^2$. From Fig.~\ref{fig:rcBK_LO} we see that no value for $C_\text{mom}^2$ can give a good agreement between the two results for all $\ktt$. In the following we fix $C_\text{mom}^2=10^3$ for which the disagreement is smaller than $30\%$ at both small and large $\ktt$. This disagreement leads to significantly different results when evaluating the NLO cross section using either Eq.~(\ref{eq:NLO_unsub}) or~(\ref{eq:NLO_sub}), as shown in Fig.~\ref{fig:rcBK_NLO_mom}. One immediately notices that the equivalence of the two equations is significantly  broken, and the ``subtracted'' cross section~\nr{eq:NLO_sub} again becomes negative, albeit at a higher scale than in the CXY prescription. This demonstrates the point emphasized in~\cite{Iancu:2016vyg} that the subtraction procedure involves a difference of two terms that are numerically, although not parametrically, large and easily becomes unstable with any further approximations. 

To solve this problem, one could in principle rewrite the expression for the cross section in coordinate space. This would allow to use a coordinate space running coupling matching the one used when solving the BK equation. However, how to implement the running coupling in a way which is consistent with the prescription~\cite{Balitsky:2006wa} is not unique, and a rather straightforward implementation leads to troublesome results (see the discussion in Appendix~\ref{sec:app_rc}). Therefore, we consider for now that the most physical prescription to use running coupling in this process is the unsubtracted result shown in Fig.~\ref{fig:rcBK_NLO_mom}.

\section{Outlook}
\label{sec:outlook}

In conclusion, we have demonstrated that it is possible to calculate single inclusive particle production in the hybrid formalism to NLO accuracy in a controlled way. While this is a promising start, many further improvements need to be made before a full phenomenological analysis and a comparison to experimental data can be made. 

Firstly it is necessary to also include the other channels in the calculation. In principle this should be straightforward. In particular there is no reason why the same rapidity factorization procedure as here could not also be used for the $g \to g$ channel, which exhibits a similar large energy logarithm that needs to be resummed using the BK equation. One also needs to add fragmentation functions to the calculation. A successful phenomenology would require better control of also these,  since so far there are large differences between different sets, in particular in LHC kinematics where the gluon channel starts to dominate~\cite{d'Enterria:2013vba}.

For a fully consistent NLO calculation one also needs to use a NLO version of the BK equation~\cite{Balitsky:2008zza,Lappi:2015fma,Lappi:2016fmu} or at least one including the resummed double and single transverse momentum logarithms~\cite{Iancu:2015vea,Iancu:2015joa} that can be made to include most of the NLO effects~\cite{Lappi:2016fmu}. Using these requires the corresponding additional term to be added to the calculation of the single inclusive cross section. One also needs to obtain the initial condition for the BK equation from a fit to DIS data in an NLO calculation~\cite{Balitsky:2010ze,Beuf:2011xd,Boussarie:2014lxa,Beuf:2016wdz,Boussarie:2016ogo}. 
Also the issue of momentum vs. coordinate space running coupling merits to be studied further. A better consistency between calculations of DIS and particle production cross sections would require a consistent running coupling scheme between the two.

The NLO corrections almost inevitably decrease the cross section from the leading order calculation. This is essentially due to the fact that, as emphasized in \cite{Stasto:2014sea,Ducloue:2016shw}, at large $\ktt$ the $\nc$-part of the unsubtracted NLO corrections behaves like $\sim \xi/\ktt^4$. The LO calculation only uses this gluon emission in the soft limit $\xi=1$ to construct the BK evolution equation, overestimating the result in exact kinematics with arbitrary $\xi$. Considering that the LO calculations such as~\cite{Lappi:2013zma} typically require ``$K$-factors'' greater than unity to make contact with the data, this might sound problematic. Here, however, it is important to keep in mind that also the NLO corrections to DIS cross sections will presumably be negative. Thus an NLO fit to DIS data could be expected to increase the normalization of the dipole amplitude compared to the current LO ones. An accurate assessment of the full phenomenological effect of the NLO corrections is impossible without a simultaneous calculation of DIS and forward pA particle production cross sections.

\section*{Acknowledgments} 
We thank E. Iancu  and D. Zaslavsky for discussions and H. Mäntysaari for providing his BK solutions. This work has been supported by the Academy of Finland, projects 267321, 273464 and 303756 and  by the European Research Council, grant
ERC-2015-CoG-681707.

\appendix

\section{Running coupling in coordinate space}
\label{sec:app_rc}

\begin{figure*}
	\includegraphics[scale=\figscale]{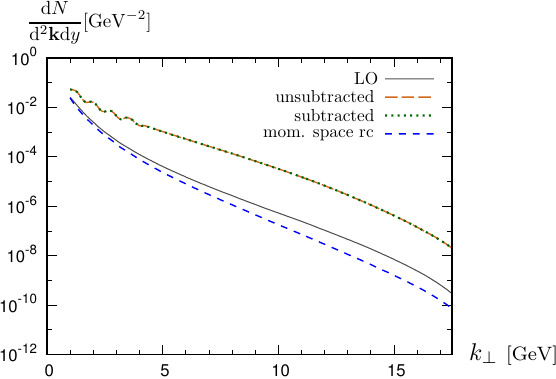}
	\hspace{\figspace}
	\includegraphics[scale=\figscale]{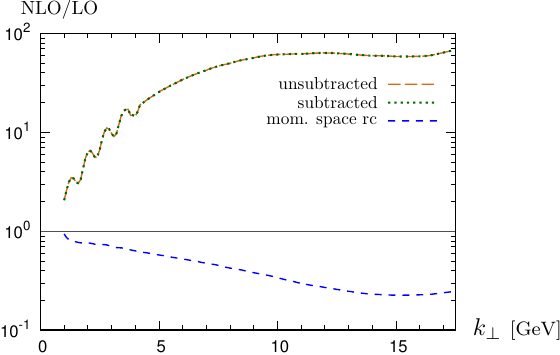}
	\caption{The complete NLO quark channel multiplicity including both the $\nc$ and $\cf$ corrections calculated with a coordinate space running coupling~(\ref{eq:as_pos}) for $\sqrt{s}=500$ GeV and at rapidity $y=3.2$, showing the NLO result using the unsubtracted~(\ref{eq:NLO_unsub}) and subtracted~(\ref{eq:NLO_sub}) expressions as well as the unsubtracted result obtained with a momentum space running coupling~(\ref{eq:as_mom}). On the left the multiplicity and on the right the ratio to the LO result.}
	\label{fig:rcBK_NLO_pos}
\end{figure*}

The equivalence between the ``unsubtracted''~(\ref{eq:NLO_unsub}) and ``subtracted''~(\ref{eq:NLO_sub}) expressions for the NLO multiplicity holds only if the coupling $\as$ used to evaluate these expressions is the same as the one used when solving the Balitsky-Kovchegov equation to obtain the rapidity dependence of the dipole correlators. In particular, as shown in Sec.~\ref{sec:rc}, this equivalence is broken if one wants to use a momentum space running coupling when evaluating the multiplicity while the prescription for the coupling used when solving the BK equation is formulated in coordinate space. In principle this problem could be solved by rewriting the expression for the multiplicity in coordinate space. A straightforward calculation shows that we can write the needed integrals $\ical^\text{finite}$, $\ical_v^\text{finite}$, $\jcal$ and $\jcal_v$ as
\begin{widetext}
\begin{align}
\label{eq:I_pos}
\ical^\text{finite}(\ktt,\xi,X(\xi)) = \int \ud^2\rt & \bigg[ \frac{S(\rt,X(\xi))}{4\pi} \ln \frac{c_{0}^{2}}{\rt^{2}\mu ^{2}}\left( e^{-i\kt\cdot \rt}+\frac{1}{\xi^{2}}e^{-i\frac{\kt}{\xi }\cdot \rt}\right) \nonumber \\
& - 2 e^{-i\kt\cdot\rt} \int\frac{\ud^2\xt}{(2\pi)^2}\frac{\xt\cdot(\xt+\rt)}{\xt^2(\xt+\rt)^2} S(\xi\rt-(1-\xi)\xt,X(\xi)) \bigg] \, ,\\
\label{eq:Iv_pos}
\ical_v^\text{finite}(\ktt,\xi,X(\xi)) = \int \ud^2 \rt & e^{-i\kt\cdot\rt} \frac{S(\rt,X(\xi))}{2\pi} \left(\ln{\frac{\ktt^2}{\mu^2}}+ \ln(1-\xi)^2\right) , \\
\label{eq:J_pos}
\jcal(\ktt,\xi,X(\xi)) = \int \ud^2 \rt & e^{-i\kt\cdot\rt} \widetilde{\jcal}(\rt,\xi,X(\xi)) \nonumber \\
=\int \ud^2 \rt & e^{-i\kt\cdot\rt} \int\frac{\ud^2\xt}{(2\pi)^2}\frac{2 \, \xt\cdot(\xt+\rt)}{\xt^2(\xt+\rt)^2}\big[ S(\rt+(1-\xi)\xt,X(\xi))-S(-\xi\xt,X(\xi))S(\rt+\xt,X(\xi))\big], \\
\label{eq:Jv_pos}
\jcal_v(\ktt,\xi,X(\xi))=\int \ud^2 & \rt e^{-i\kt\cdot\rt} \widetilde{\jcal}_v(\rt,\xi,X(\xi)) \nonumber \\
=\int \ud^2 \rt & e^{-i\kt\cdot\rt} \int\frac{\ud^2\xt}{(2\pi)^2}\frac{2}{\xt^2}\big[S(\rt-(1-\xi)\xt,X(\xi))
-S(-\xt,X(\xi))S(\rt+\xi\xt,X(\xi))\big] .
\end{align}
\end{widetext}
With these coordinate space expressions it is possible to move the coupling inside the integrals and replace the kernels by an expression that (a) reduces to the same expressions for a fixed coupling and (b) reduces to a desired coordinate space running coupling BK kernel in the limit $\xi=1$. These two requirements do naturally not uniquely determine the expression at other values of $\xi$. In the notations introduced above the BK equation can be written as
\begin{widetext}
\begin{equation}
\partial_Y S(\rt,X)= 2\as\nc\left[\widetilde{\jcal}(\rt,1,X)-\widetilde{\jcal}_v(\rt,1,X)\right]
=2\as\nc\int\frac{\ud^2\xt}{(2\pi)^2}\frac{\rt^2}{\xt^2(\xt+\rt)^2} \big[S(-\xt,X)S(\rt+\xt,X)-S(\rt,X)\big] ,
\end{equation}
at the rapidity $Y=\ln\frac{1}{X}$. The Balitsky prescription~\cite{Balitsky:2006wa} for the running coupling consists in making the replacement $\as \to \as(\rt^2)$ and in a modification of the kernel, leading to
\begin{align}
\partial_Y S(\rt,X)=2\alpha_s(\rt^2)\nc\int\frac{\ud^2\xt}{(2\pi)^2}&\big[S(-\xt,X)S(\rt+\xt,X(\xi))-S(\rt,X)\big] \nonumber \\
&\times\bigg[\frac{\rt^2}{\xt^2(\xt+\rt)^2}+\frac{1}{\xt^2}\left(\frac{\alpha_s(\xt^2)}{\alpha_s((\xt+\rt)^2)}-1\right)+\frac{1}{(\xt+\rt)^2}\left(\frac{\alpha_s((\xt+\rt)^2)}{\alpha_s(\xt^2)}-1\right)\bigg] .
\end{align}
\end{widetext}
Therefore we find that a rather straightforward way to implement this prescription for arbitrary $\xi$ is to replace $\widetilde{\jcal}_v$ with
\begin{align}
&\widetilde{\jcal}_v^\text{rc}(\rt,\xi,X) =\int\frac{\ud^2\xt}{(2\pi)^2}\frac{2}{\xt^2}\frac{\alpha_s(\xt^2)}{\alpha_s((\xi\xt+\rt)^2)} \nonumber \\
&\times\big[S(\rt-(1-\xi)\xt,X)
-S(-\xt,X)S(\rt+\xi\xt,X)\big]\, ,
\end{align}
as well as to move the coupling $\as$ appearing in Eqs.~(\ref{eq:nc_unsub}), (\ref{eq:nc_sub}) and~(\ref{eq:cf}) inside the $\rt$-integrals~(\ref{eq:I_pos})-(\ref{eq:Jv_pos}), then making the replacement $\as \to \as(\rt^2)$. It is easy to check that the two conditions (a) and (b) mentioned previously are satisfied with this choice, which however is not unique.
The numerical implementation of the ``unsubtracted''~(\ref{eq:NLO_unsub}) and ``subtracted''~(\ref{eq:NLO_sub}) expressions for the NLO multiplicity using this coordinate space formulation is shown in Fig.~\ref{fig:rcBK_NLO_pos}. This figure demonstrates that, contrary to the results obtained with a momentum space running coupling shown in Fig.~\ref{fig:rcBK_NLO_mom}, the unsubtracted and subtracted results are the same, meaning that our modification of the $\nc$-terms matches the Balitsky prescription at $\xi=1$. However, these results are problematic since the behavior of the NLO multiplicity is totally different from the results obtained at fixed coupling or with a momentum space running coupling, with a NLO result orders of magnitude larger than the LO one. This issue may be related to the fact that our implementation of the coordinate space running coupling, while quite straightforward, is not unique and other choices could give different results. How to consistently generalize the Balitsky prescription to $\xi \ne 1$ is a non-trivial issue that goes beyond the scope of this work.

\bibliography{spires}
\bibliographystyle{JHEP-2modlong}
\end{document}